\documentclass[twocolumn,aps,pra,showpacs,superscriptaddress,floatfix,citesort]{revtex4}
\usepackage{epsfig}  

\newcommand{\gl}[1]{Eq. (\ref{#1})}

\def\gtrless{\raise2.5pt\hbox{$>$}\llap{\lower2.5pt\hbox{$<$}}}
\def\gtrapprox{\raise2.5pt\hbox{$>$}\llap{\lower2.5pt\hbox{$\approx$}}}
\newcommand{\bsq}[1]{\begin{subequations}\label{#1}}
\newcommand{\esq}{\end{subequations}}
\newcommand{\beq}[1]{\begin{equation}\label{#1}}
\newcommand{\eeq}{\end{equation}}
\newcommand{\beqa}[1]{\begin{eqnarray}\label{#1}}
\newcommand{\eeqa}{\end{eqnarray}} \newcommand{\fur}{\qquad\mbox{for }\, } 
\newcommand{\wer}{\qquad\mbox{where }\quad}
 
\newcommand{\gd}{\dot{\gamma}}

\newcommand{\eps}{\varepsilon}
\renewcommand{\rho}{\varrho}

\newcommand{\rem}[1]{}

\begin{document}

\title{Nonlinear rheology of dense colloidal dispersions: 
a phenomenological model and its connection to mode coupling theory}

\author{Matthias Fuchs}
 \email[Corresponding author. Email: ]
  {matthias.fuchs@uni-konstanz.de}
\affiliation{Fachbereich Physik, Universit\"at Konstanz,
 78457 Konstanz, Germany
}

\author{ Matthias Ballauff}
\affiliation{Physikalische Chemie I, Universit\"at Bayreuth,
95440 Bayreuth, Germany}

\date{\today}


\begin{abstract}
Rheological properties, especially 'shear--thinning',
 of dense colloidal dispersions are discussed on
three different levels. A generalized phenomonological Maxwell model
gives a broad framework connecting glassy dynamics to the linear and
non--linear rheology of dense amorphous particle
solutions. First principles mode coupling theory calculations for the time
or frequency dependent shear modulus give  quantitative results for
dispersions of hard colloidal spheres in the linear regime. Schematic
models extending mode coupling theory to the non--linear regime
recover the phenomenology of the generalized Maxwell model, and 
predict universal features of flow curves, stress versus shear--rate.
\end{abstract}
\maketitle

\section{Introduction}

The rheological behavior of 
dispersions of rigid colloidal particles is immensely rich
\cite{larson,russel}.  Universally they exhibit shear--thinning, that
the viscosity $\eta$ decreases from the Newtonian--plateau upon
increasing the shear rate $\gd$.
Another phenomenon in dense colloidal dispersions, or dispersions of
particles with strong attractions, is that they undergo
arrest--transitions and solidify to (mostly weak) amorphous solids (glasses)
\cite{Megen93,Megen93b,Megen94,Beck99,Bartsch02,Eckert03,Mason95}.
In this contribution, we want to establish under which circumstances and
simplifications one may expect that the physics of the glass transition
determines or even dominates the rheological behavior of colloidal
dispersions. Our discussion supplements a recent comparison of theoretical computations
and measurements of flow curves in colloidal model systems where 
this connection is explored quantitatively \cite{Fuchs04}.

While recently microscopic approaches to shear--thinning in colloidal
dispersions  have been presented \cite{Fuchs02,Fuchs03b,Fuchs03,Cates04}, in
this contribution we first argue on a more qualitative level and 
sketch  a phenomenological picture, that goes back to Maxwell. Then,
we discuss in how far simple 'schematic models' derived from the 
first principles approaches in Refs.
\cite{Fuchs02,Fuchs03b,Fuchs03,Cates04},
support the generalized phenomenological Maxwell model.
The concepts we discuss build upon insights into the structural
dynamics obtained with the mode coupling theory (MCT) of G\"otze and coworkers
\cite{Goetze91b,gs}, and thus extend a microscopic 
approach which has amply been tested and compared to computer
simulation studies and experimental data from quiescent
model systems \cite{Goetze99}. We also present MCT predictions for the linear
viscoelasticity of colloidal hard spheres which may be tested in future simulational or
rheological studies.

\section{A phenomenological generalized Maxwell model}

\subsection{Quiescent system}

Before discussing the connection of glassy dynamics with the rheology
of dense colloidal dispersions, it appears helpful to
reconsider the oldest phenomenological description of the glass
transition which goes back to Maxwell. Maxwell considered the most
general constitutive equation relating (transversal) stress $\sigma$ and
shear rate 
$\gd$ within the framework of linear response theory:
\beq{maxwell1}
\sigma(t) = \int_{-\infty}^{t}\!\!\!\!dt'\; 
G^{\rm tot}(t-t')\; \gd(t') \; .
\eeq
Here, $G^{\rm tot}(t)$ is the shear modulus and is, within linear response, 
independent of the applied  shear rate $\gd$, which in the most
general situation is a time--dependent function.  This relation  is
exact for the linear response of systems with free energy corresponding
to a fluid phase provided that the system was in equilibrium 
before the (smoothly switched--on) application of shear
\cite{Forster75}.
 For colloidal solutions, Batchelor 
was the first to discuss the three contributions to the stress which
arise: in the solvent, from the potential interactions among the
particles, and from the solvent particle interactions
\cite{Batchelor77}.
 For the
dispersions of interest here, the solvent can be considered a
continuum fluid characterized by a  know solvent viscosity $\eta_{s}$,
so that $G^{\rm solv}(t)=\eta_{s}\delta(t)$.   

Solvent--particle interactions (viz. the HI) act instantaneously if
the particle microstructure differs from the equilibrium one, but do
not themselves determine the equilibrium structure.  If one assumes
that glassy arrest is connected with the ability of the system to
explore its configuration space and to approach its equilibrium
structure, then
 it appears natural to assume that the solvent particle interactions 
are characterized by a finite  time scale
$\tau_{\rm HI}$.  In a most naive way, the contribution from HI,
$G^{\rm HI}(t)$,
 to the shear modulus $G^{\rm tot}(t)$ would thus be written as:
$G^{\rm HI}(t) = G^{\rm HI}_{\infty} \exp{\{-t/ \tau_{\rm HI}\}}$,
where the time $\tau_{\rm HI}$ stays finite and varies little with the
distance to glassy arrest. Considering time--independent shear rates,
$\gd(t)=\gd$, so that
\beq{maxwell1h}
\sigma = \eta \; \gd = \gd \; \int_{0}^{\infty}\!\!\! dt \; G^{\rm
  tot}(t)\; ,
\eeq
HI would thus lead to an increase of the high frequency
viscosity above the solvent
value; this value shall be denoted as
$\eta_{\infty} = \eta_{s}+ G^{\rm HI}_{\infty} \,\tau_{\rm HI}$.

Maxwell in his studies of viscous liquids considered the potential
contribution, to be abbreviated as $G(t)$, to the shear modulus, which
arises from the forces between the particles. Its microscopic definition is,
$G(t)=\langle \sigma_{xy}(t)
\sigma_{xy}(0)\rangle/(k_{B}TV)$, where the relevant element of the
fluctuating stress tensor is given with the potential forces ${\bf
  F}_{i}$ and particle coordinates ${\bf r}_{i}$ 
by $\sigma_{xy}(t)= - \sum_{i}
F^{x}_{i}(t) \,y_{i}(t)$. Kinetic
contributions that would differ between Newtonian (which Maxwell
considered) and Brownian (the colloidal dispersions considered here) systems 
do not contribute for symmetry reasons. He postulated that close to
the glass transition the modulus develops a slow relaxational process
that is characterized by a time scale $\tau$, which grows beyond
bounds:
\beq{maxwell2}
G(t) = G_{\infty} \; \exp{\{- \frac t\tau\}} \; .
\eeq
The amplitude of the slow relaxational process, $G_{\infty}$, can be
considered an almost constant that changes much more slowly with
density or other parameters than the time scale $\tau$. 
The divergence of $\tau$ upon solidification often is modeled  \cite{larson} by
 a Krieger--Dougherty, $\tau/\tau_{0}\propto (1-\phi/\phi_{m})^{-2}$,
or a Doolitle form,  $\ln{\tau/\tau_{0}} \propto
(1-\phi/\phi_{m})^{-1}$.
Here the packing fraction at the maximal fluid density is denoted
as $\phi_{m}$, and the time--unit $\tau_{0}$ should be connected to the
viscosity $\eta_{\infty}$ that a particle feels because of the
surrounding solvent. The slow process described by Maxwell, which in the
glass literature is called $\alpha$--process, causes a marked and
strongly density dependent increase
of the total viscosity over $\eta_{\infty}$:  performing the
integration gives
$\eta_{0}= \eta_{\infty } + G_{\infty} \; \tau$. In this picture,
the total shear modulus as a function of time consists of the three
mentioned terms: $G^{\rm tot}(t)=\eta^{s}\delta(t)+G^{\rm
  HI}(t)+G(t)$. 

The naive picture sketched here, is not correct for a number of
reasons. It is well known that for hard spheres without HI the shear
modulus diverges for short times, $G^{\rm HS no HI}(t\to0)\sim
t^{-1/2}$. Lubrication forces, which keep the particles apart,
  eliminate this divergence and render
$G^{\rm tot}(t\to0)$ finite \cite{Lionberger94}.
 Thus, the simple separation of $G^{\rm
  HI}(t)$ and the potential $G(t)$ is not possible for short times, at
least for particles with a hard core. Moreover, comparison of
simulations without and with HI has shown that the
increase of $(\eta_{0}-\eta_{\infty})/\eta_{\infty}$ depends somewhat
 on HI, and thus not just on the potential interactions as implied.

  Nevertheless the sketched picture provides the most basic
view of a glass transition in colloidal suspensions, 
connecting it with the increase of the structural relaxation time $\tau$. 
Increased density or interactions cause a slowing down of particle
rearrangements which leave the HI relatively unaffected, as these
solvent mediated forces act on all time scales. Potential forces
dominate the slowest particle rearrangements because vitrification
corresponds to the limit where they actually prevent the final
relaxation of the microstructure. The structural relaxation time
diverges at the glass transition, while $\tau^{\rm HI}$ stays
finite. Thus close to arrest a time scale separation is possible,
$\tau\gg\tau^{{\rm HI}}$.

While the parameter
$G_{\infty}$ in Maxwell's model might appear to play a rather boring role as
amplitude of the final relaxation process, its actual significance is
quite high. It gives the elastic constant of the arrested structure in
the limit of $1/\tau=0$ (viz. in the glass),
 or of the glassy structure present on
intermediate times, $t\ll \tau$.  Thus in the glass or for shear rates
$\gd(t)$ that vary more rapidly with time than the modulus, a Hookian elastic
response follows from \gl{maxwell1}:
\beqa{maxwell3}
\sigma(t) &=& G_{\infty} u(t) + \mbox{dissipative terms ,} \wer \\
u(t) &=& \int_{-\infty}^{t}\!\!\!dt' \gd(t')\fur{1/\tau=0} \; ,
\eeqa
where $u(t)$ is the strain imposed on the amorphous solid. 
Clearly, $G_{\infty}$ plays the role of a spring constant for the
elastic restoring forces in the solid which grow with the imposed
distortion measured by the strain, viz. $u(t) = \int^{t}\!\!dt'\;
\partial v_{x}(t')/\partial y$.

An
interesting difference from this hall mark behavior of glasses to
colloidal coagulation or gelation at low packing fractions
should be noted. Experiments on a wide range of
gels at high dilutions have shown that the elastic
constant $G_{\infty}$ vanishes at the colloidal gelation line; it then
grows with a strong power--law upon entering deeper into the gel
phase \cite{Trappe00}. 
It has been pointed out that this behavior can be explained by
an underlying percolation like transition.  
 Thus the observed 'Trappe--scaling' points to a deep difference
between colloidal coagulation at very low packing fractions and
colloidal vitrification at higher ones. At the latter transition,
$G_{\infty}$ is finite, as first postulated by Maxwell. 
Interestingly, this difference
does not show up in universal jamming phase diagrams.
Intriguing data have been obtained for 
intermediate concentration range where both  pictures transform, but
further studies appear very fruitful \cite{Prasad03}.

The elastic constant $G_{\infty}$   serves to indicate a difference of
the present Maxwell's view on the glass transition to another colloid
theory. It is the approach by Brady \cite{Brady93}, who
 identifies the solidification of amorphous  dispersions 
with random close packing, $\phi_{m}=\phi_{\rm rcp}$. 
In this approach, the modulus $G^{\rm tot}$ relaxes in a single 
relaxational process characterized by the time scale $\tau^{\rm
  s}=a^{2}/ D^{s}_{0}$; $a$ is the particle radius, and $D^{s}_{0}$
the density dependent short--time self diffusion constant. While
Brady's  final result for the viscosity, $\eta_{0}\propto
(1-\phi/\phi_{m})^{-2}$, is  similar to
glassy behavior (MCT predicts a divergence $\eta_{0}\propto\tau\propto
(1-\phi/\phi_{c})^{-2.6}$ for hard spheres), part of the divergence of
$\eta_{0}$ comes from a divergence of the elastic constant upon
approaching random close packing, $G_{\infty} \propto
(1-\phi/\phi_{m})^{-1}$. The time scale $\tau^{\rm s}$ also diverges
at solidification, but contributes only comparable to the amplitude of
the relaxational process,  $\tau^{\rm  s} \propto (1-\phi/\phi_{m})^{-1}$.
Thus, a quite different scenario of arrest at random close packing 
to arrest at the glass transition should become apparent if $G^{\rm
tot}(t)$ or its Fourier--transforms, elastic $G'(\omega)$ and loss
$G''(\omega)$ moduli are measured. (We use the convention
$G''(\omega)+iG'(\omega) = \omega \int_{0}^{\infty}\!\!\!dt\, e^{i \omega
  t}\, G(t)$) While at the glass transition the dynamics
should only become slower, at random close packing the amplitude of
the stress at intermediate times should also increase. Besides this
qualitative difference, a quantitative difference also arises in the
values of arrest, namely the glass transition density, called
$\phi_{c}$ in MCT and measured in, lies below random close packing:
$\phi_{c}\approx 0.58<\phi_{rcp}\approx 0.64$
\cite{Megen93,Megen93b,Megen94,Beck99,Bartsch02,Eckert03}.  Particles
in a glass 
thus still have local volume to rearrange; they can explore their
respective 'cages', but cannot relax the overall microstructure.

\subsection{Under shear}

The utility of the phenomenological Maxwell model of the glass
transition is further stressed by the ease by which it can be extended to
the shear--thinning of colloidal fluids and yielding of colloidal
glasses  under shear. The central concept required is that shearing
speeds up the relaxation of fluctuations and kills off the long--time
memory present in very viscous fluids. This requires that shear causes
the slow structural or $\alpha$--relaxational process in $G(t)$ of the
Maxwell model to decay faster. It is natural to connect the relaxation
time of the structural relaxation to the inverse of the shear rate,
and thus to formulate on the phenomenological level:
\beq{maxwell4}
g^{\rm gMm}(t,\gd) = G^{\rm tot}(t) \; \exp{\{- t \; |\, \gd\, | \}}\; .
\eeq
Here, the generalized non--linear shear modulus $g(t,\gd)$ is a
function describing the transient fluctuations in a dispersion sheared
from time $t=0$ on.
Equation (\ref{maxwell4}) expresses that under shear transient
stress fluctuations either decay via the relaxation channels present
without shear, or via a shear induced relaxation mechanism whose
intrinsic decay rate is set by the shear rate itself.
Postulating that the shear rate enters in a non--analytic way via the
absolute value into the phenomenological description is of course, a
priori, questionable, and requires the  derivation given in
Refs. \cite{Fuchs02,Fuchs03}; see also section IV. 
The result \gl{maxwell4} is quite satisfying, because, intuitively, one
expects the direction of the solvent flow profile to be irrelevant for
the magnitude of the stress or viscosity even in the non--linear regime. 

The generalization of \gl{maxwell1}, that connects $g(t,\gd)$ to the
steady state stress in the case of constant shearing is:
\beqa{maxwell5}
\sigma(\gd) &=& \eta(\gd)\; \gd = \gd \; \int_{0}^{\infty}\!\!\!dt\;
g^{\rm gMm}(t,\gd) \nonumber \\
&=& \gd\; \left\{ \eta_{s} + \frac{\eta_{\infty}-\eta_{s}}{1+|\gd| \;
\tau_{{\rm HI}}} + \frac{\eta_{0}-\eta_{\infty}}{1+|\gd| \;
\tau} \right\} \; .
\eeqa 
This is the final expression of the extended Maxwell model for the stationary
viscosity or stress as a function of the shear rate, $\eta(\gd)$.
Variations of the packing fraction or of other control parameters of the
system without shear enter via the Maxwell
relaxation time $\tau$, and consequently $\eta_{0}$, which may for
example be modeled by a Krieger 
Dougherty expression below the glass transition density, and by
$\tau=\infty=\eta_{0}$ above it.  All other constants appearing
explicitly or implicitly in \gl{maxwell5} may, in the most simple
approach, be taken as control--parameter independent constants.

The discussed
phenomenology is quite rich with flow curves that exhibit $(i)$ a
Newtonian viscosity plateau given by $\eta_{0}$ at low shear rates and
below the glass transition (for $|\gd|\ll1/\tau\ll1/\tau_{\rm HI}$), 
$(ii)$ a shear thinning region for $1/\tau
\ll |\gd| \ll 1/\tau_{{\rm HI}}$ below the glass transition,
 $(iii)$ a solid--like yield flow  for $0=1/\tau \ll |\gd| \ll
1/\tau_{{\rm HI}}$ above the glass transition, and $(iv)$ a second
Newtonian plateau characterized by $\eta_{\infty}$ following the shear
thinning region while still  $|\gd| \ll 1/\tau_{\rm HI}$ holds. 
In this region of high shear, where the HI are dominant, the suggested
modeling of $G^{\rm HI}(t)$ is not based on theoretical insights from
our microscopic
 approach, and thus is not reliable. Therefore, the most sensible
approach, that we also will follow in the remainder of this presentation,
is to restrict the shear rates to the region $|\gd| \tau_{\rm HI}\ll
1$. Then \gl{maxwell5} simplifies to 
$\eta(\gd)=\eta_{\infty} + \left( \frac{1}{\eta_{0}}
+\frac{|\gd|}{\sigma^{*}} \right)^{-1}$,  which
captures the small shear rate limit close to vitrification, where
$\tau\gg\tau_{\rm HI}$, and $\eta_{0}\gg\eta_{\infty}$.
The notation $\sigma^{{*}}$ has been introduced, to stress the
dimensionalities; from \gl{maxwell5} $\sigma^{{*}}=G_{\infty}$
follows, while in more general settings $\sigma^{*}$ is more closely
related to the critical yield stress at solidification \cite{Cates04}. 

Aim of section IV is to present and discuss a schematic
model, which can be derived within our microscopic approach
\cite{Fuchs02,Fuchs03},  and which
recovers the phenomenology just described.
 
\section{Viscoelasticity of quiescent colloidal fluids}

Mode coupling theory (MCT) appears to capture the cage--effect
and predicts that it dominates the slow relaxation of
structural correlations \cite{gs}. Density fluctuations play an important role 
because they are well suited to describe the structure
of the particle system and its relaxation. Moreover, stresses that
decay slowly because of the slow particle rearrangements, MCT argues,
also  can be approximated by density fluctuations
 using effective potentials. Density fluctuations not at large
wavelengths, but for wavelengths corresponding to the average particle
distance turn out to be the dominant ones. In agreement with the picture
of the caging of particles by structural correlations, the MCT
glass transition is independent on whether the particles move ballistically
in between interactions with their neighbors (say collisions for hard
spheres) or by diffusion. Structural arrest happens whenever the
static density correlations for wavelengths around the average
particle distance are strong enough. Thus, it had been realized early
on that this basic version of MCT, the 'idealized 
MCT', that was developed for a description of the structural arrest in
atomic systems also needs to apply to colloidal dispersions \cite{gs}. 
\begin{figure}[h]
\centering
\includegraphics[width=0.4\textwidth]{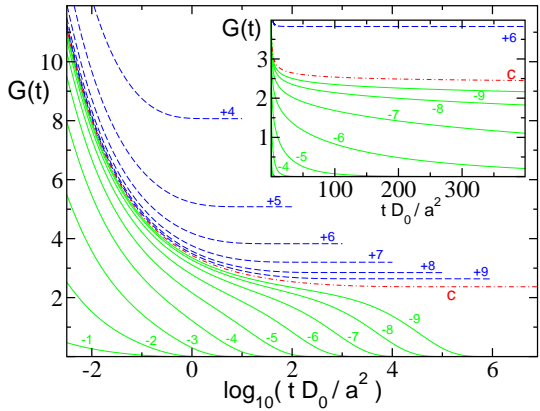}
\caption{\label{bild2} Equilibrium potential shear modulus
  $G(t)$ (in units of $k_{B}T/a^{3}$) for colloidal
  hard spheres  with radius $a$ for packing fractions close to
  solidification at $\phi_{c}$. Densities are measured by the separation
  parameter $\varepsilon=(\phi-\phi_{c})/\phi_{c}=\pm 10^{-|n|/3}$, and labels
  denote the value $n$. Positive values belong to glass ($\eps>0$),
  negative to   fluid states ($\eps<0$);
 the label $c$ gives the transition. The inset shows a
  subset of the curves on a linear time axis; the increase of $G(t)$
  for short times cannot be resolved. }
\end{figure}

Results of the (idealized) MCT equations for the potential part 
of the
time--dependent shear
modulus $G(t)$ of hard spheres for various packing fractions $\phi$
are shown in Fig. \ref{bild2} and calculated from \cite{Goetze91b}:   
\beq{mi1}
G(t)  \approx
\frac{k_BT}{60\pi^2} \;
\int_0^\infty\!\!dk\;  k^4\; \left(
\frac{\partial \ln{S_k}}{\partial k} \right)^2\; \Phi^2_{{ k}}(t)\; ,
\eeq
The normalized density fluctuation functions $\Phi_{k}(t)=\langle
\delta\rho^{*}({\bf k},t)\delta\rho({\bf k})\rangle/\langle|
\delta\rho({\bf k})|^{2}\rangle/$ enter, which are calculated self--consistently 
within MCT \cite{Franosch98}.  Packing fractions are measured in
relative separations $\eps=(\phi-\phi_{c})/\phi_{c}$ to the
glass transition point, which for this model of hard 
spheres lies at $\phi_{c}=0.516$ \cite{Goetze91b}.
 Note that this result depends on the
static structure factor $S(q)$ only, which is taken from
Percus--Yevick theory, and that the experimentally determined
value $\phi_{c}^{\rm expt.}=0.58$ lies somewhat higher \cite{Megen93,Megen94}.
For low packing fractions, or large negative separations, the modulus
decays quickly on a time--scale set by the short--time diffusion of
well separated particles. The strength of the modulus increases
strongly at these low densities, and its behavior at short times
presumably 
depends sensitively on the details of hydrodynamic and potential
interactions; thus Fig. \ref{bild2} is not continued to small times,
where the employed model (taken from Ref. \cite{fuchsmayr}) is too
crude.   For smaller (negative) separations from the critical density,
$\eps\nearrow0$, little changes in $G(t)$ at short times, because the
absolute change in density becomes small. Yet, at long times a process
in $G(t)$ becomes progressively slower upon taking $\eps$ to zero. It
can be considered the MCT analog of the phenomenological
Maxwell--process.
MCT finds that it depends on the equilibrium structural correlations
only, while HI and other short time effects only shift its overall
time scale.  Importantly, this overall time scale
applies to the slow process in coherent and 
incoherent density fluctuations as well as in the stress
fluctuations \cite{Franosch98}. This holds even though 
e.g. HI are known to affect short time
diffusion coefficients and 
high frequency viscosities differently. 
 Upon crossing the glass
transition bifurcation, the relaxation freezes out and the amplitude
$G_{\infty}$ of the Maxwell--process does not decay; the modulus for
long times approaches the elastic constant of the glass
$G(t\to\infty)\to G_{\infty}>0$. Entering deeper into the glassy phase
the elastic constants increase quickly with packing fraction.  
\begin{figure}[h]
\centering
\includegraphics[width=0.4\textwidth]{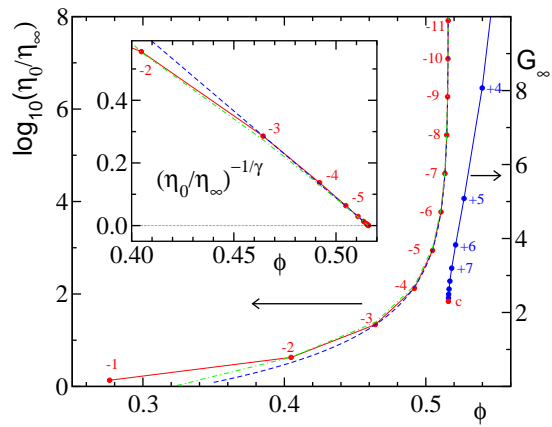}
\caption{\label{bild2h} On the left side, 
increase of the Newtonian low frequency
shear viscosity $\eta_0$ relative to the high frequency one, $\eta_\infty$,
for hard spheres as function of the packing
fraction $\phi$; numbers (negative as $\phi<\phi_{c}$) denote the separation
$\varepsilon=(\phi-\phi_{c})/\phi_{c}=\pm 10^{-|n|/3}$. 
Neglecting hydrodynamic interactions, 
$\eta_\infty=\frac{k_BT}{6\pi D_0a}$ is obtained from the short time
diffusion coefficient. The dashed line gives the asymptotic power law
$(\phi_c-\phi)^\gamma$ predicted by MCT with $\gamma=2.46$, while the 
relaxation times $\tau_\eta$ obtained from $G(\tau_\eta)= k_BT/(2a)^3$
are multiplied by a factor $0.2^{-\gamma}$ and
shown by a dot--dashed line. The inset presents the same data in a
rectification plot, $(\eta_{0}/\eta_{\infty})^{{-1/\gamma}}$, where the
power--law becomes a straight line.\newline
On the right side, the long time
limits of the shear modulus,
$G(t\to\infty)=G_{\infty}$ in units of $k_{B}T/a^{3}$,
are shown versus $\phi$, with positive
numbers $n$ denoting the separation from $\phi_{c}$.}
\end{figure}

Calculating the viscosity according to \gl{maxwell1h},
$\eta_{0}=\eta_{\infty}+\int_{0}^{\infty}\!\!\!dt\, G(t)$ it is important
to realize that the integration  is dominated by the
slow process.  The inset of Fig. \ref{bild2} highlights this by
showing $G(t)$ on a linear axis of time. The viscosity increases as
the area under $G(t)$, which is almost exclusively set by the slowest
process, which is the Maxwell one. The rapid decrease
of $G(t)$ to the plateau value $G_{\infty}$ does not affect the
viscosity as long as time--scale separation holds and  the Maxwell
process is much slower. Figure \ref{bild2h} exhibits the obtained
increase of the low frequency viscosity $\eta_0$ for a quiescent
colloidal dispersion of hard spheres. Also shown is the asymptotic power
law predicted by MCT (with prefactor fitted), 
and the variation of the longest relaxation time 
$\tau_\eta$ of $G(t)$ estimated by $G(\tau_\eta)= k_BT/(2a)^3$.  Asymptotically
all three quantities exhibit the same dependence on $\phi$, and diverge at
the glass transition density $\phi_c$. The asymptotic power law variation 
holds for relative separations up to around $-\eps\approx $10 \%.  
\begin{figure}[h]
\centering
\includegraphics[width=0.4\textwidth]{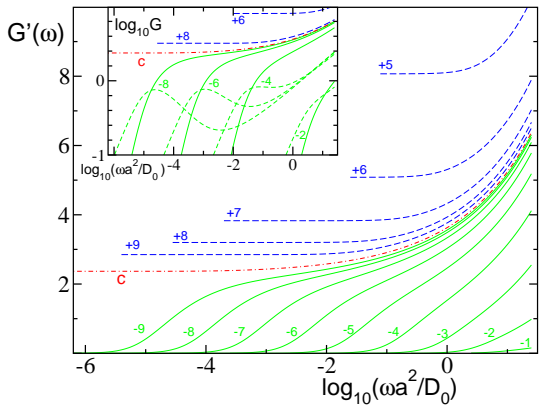}
\caption{\label{bild3} Storage part of the
shear modulus $G'(\omega)$ corresponding to 
Fig. \protect\ref{bild2}. The inset shows storage and loss moduli (only for
fluid states) for a number of densities. }
\end{figure}
\begin{figure}[h]
\centering
\includegraphics[width=0.4\textwidth]{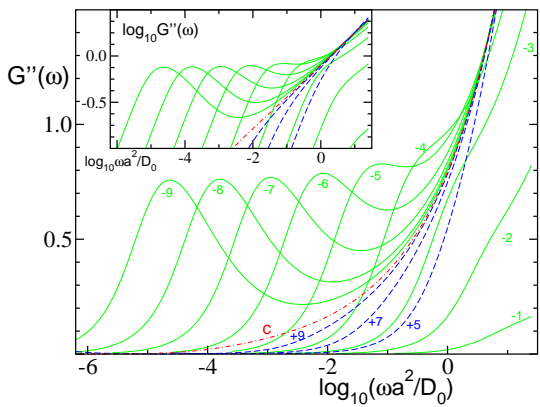}
\caption{\label{bild4} Loss part of the shear modulus 
$G''(\omega)$ corresponding to 
Fig. \protect\ref{bild2}. The inset shows the same data in a double 
logarithmic representation.}
\end{figure}

MCT thus provides a microscopic description of the linear viscoelasticity
that follows the phenomenology first presented by Maxwell; of course
in detail important and interesting differences exist, e.g. the exponential
shape of the final decay postulated by Maxwell is replaced by a
stretched--exponential decay in MCT. Tests of these specific
predictions have greatly contributed to the verification of MCT
\cite{Goetze99}. 

While $G(t)$ can be accessed by theory more directly, experimentally
its Fourier--transforms can be studied more easily. Storage
$G'(\omega)$ and loss $G''(\omega)$ modulus are shown as functions of
frequency in
Figs. \ref{bild3} and \ref{bild4} respectively. The slow
Maxwell--process appears as a shoulder in $G'$ which extends down to
lower and lower frequencies when approaching glassy arrest, and
reaches to zero frequency  in the glass, $G'(\omega=0)=G_{\infty}$. 
The slow process shows up as a peak in $G''$ which in parallel motion
(see inset of Fig. \ref{bild3}) shifts to lower frequencies when
$\eps\nearrow0$.
Including, in the spirit of the discussion in Sect.II,
 hydrodynamic interactions into the calculation would affect
the frequency dependent moduli at frequencies satisfying
$\omega\tau_{\rm HI}\approx 1$. For smaller frequencies, only a small
correction would arise,
$G''_{\rm HI}(\omega\ll1/\tau_{\rm HI})\sim \omega \eta_{\infty}$
and  $G'_{\rm HI}(\omega\ll 1/\tau_{\rm HI})\sim (\omega
\eta_{\infty})^{2}/G^{\rm HI}_{\infty}$, as long as time scale
separation and consequently $\eta_{0}\gg \eta_{\infty}$ holds.

\section{Schematic models for the non--linear rheology}

While within the MCT the linear viscoelastic regime has been worked out
quantitatively, the recent generalization of MCT to the non--linear regime
\cite{Fuchs02,Fuchs03} has not yet been fully solved.
Only a number of universal predictions have been obtained, which
exist in any model that exhibits the generic
bifurcation scenario from yielding solid to shear thinning fluid. In this section
we verify that this scenario recovers the phenomenological Maxwell picture discussed
in Sect. II. Hereto we use the most simple schematic model that has also recently been
 successfully employed to analyse flow curves from simulations \cite{Fuchs03b,Cates04b} and
 experiments \cite{Fuchs04}.

\begin{figure}[h]
\centering
\includegraphics[width=0.4\textwidth]{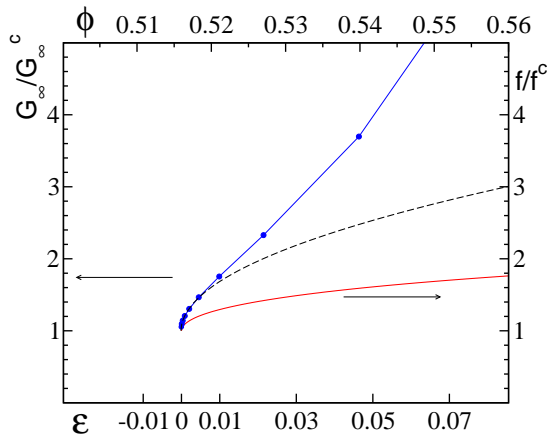}
\caption{\label{bild6} 
Normalized frozen--in glass parameters for a microscopic model, viz
the elastic constant of a hard sphere glass $G_{\infty}/G_{\infty}^{c}$, 
and a schematic model, viz. the non--ergodicity parameter $f/f_{c}$ of
the $F_{12}$--model, as functions of the separation parameter $\eps$;
the normalization is the respective value at the transition. The upper
axis gives $\eps$ expressed in packing fractions for hard spheres.}
\end{figure}

The central feature of the  equations of motion in Refs. \cite{Fuchs02,Fuchs03} is that
they contain the competition of two effects. First, a non--linear
memory effect  increases with increasing particle
interactions (`collisions' or `cage effect') which leads to a
non--ergodicity transition in the absence of shear, and second, memory
effects vanish with time because of shear-induced decorrelation. Both these
effects can be captured in the simpler `schematic'
models also. Importantly, the models can be made to obey
similar stability equations as the microscopic approach, and thus 
the universal predictions hold.

The well studied and comparatively simple schematic
F$_{12}^{(\gd)}$--model considers one normalized correlator $\Phi(t)$,
 which obeys a generalized relaxation equation:
\beq{sm1}
\dot{\Phi}(t) + \Gamma  \left\{
\Phi(t) +  \int_0^t\!\!\! dt'\; m(t-t') \, \dot{\Phi}(t')
\right\} = 0 \; .
\eeq
Without memory effects, $m\equiv0$, the correlator relaxes
exponentially, $\Phi(t)=\exp{-\Gamma t}$, but with $m\neq 0$,
retardation effects set in after a short--time variation (still given
by the initial decay rate $\Gamma$, viz.  $\Phi(t\to0)=1-\Gamma t+\ldots$). 
The correlator $\Phi(t)$ may be thought to correspond to the
normalized shear modulus of the generalized Maxwell model of section II.
A low order polynomial ansatz for $m$ suffices to model the feedback
 mechanism of the cage--effect.  We choose
\beq{sm2}
 m(t) = \frac{1}{1+(\gd t)^2} \; 
 \left(v_1 \Phi(t)+v_2 \Phi^2(t) \right) \; .
\eeq
Without shear, this model has been studied extensively
\cite{Goetze84,Goetze91b}. Increasing particle caging is modeled by
increasing coupling parameters $v_1$, $v_2 \ge0$, and the only effect of
shearing is to cause a time dependent decay of the friction kernel $m$.
The system loses memory because of shearing.
 The role of the transport coefficient
(viscosity) $\eta$ is played by the average relaxation time obtained
from integrating the correlator, and this also is taken to determine
the stress: 
\beq{sm3}
\sigma = \gd\;  \eta = 
\gd \; \langle \tau \rangle = \gd \int_0^\infty\!\!\! dt\;  \Phi(t) \; .
\eeq
At high shear rates, the memory function is strongly supressed, so that
$\Phi$ returns to a single 
exponential, and the high shear visosity of the model
follows as $\eta_{\infty}=1/\Gamma$.

\begin{figure}[h]
\centering
\includegraphics[width=0.4\textwidth]{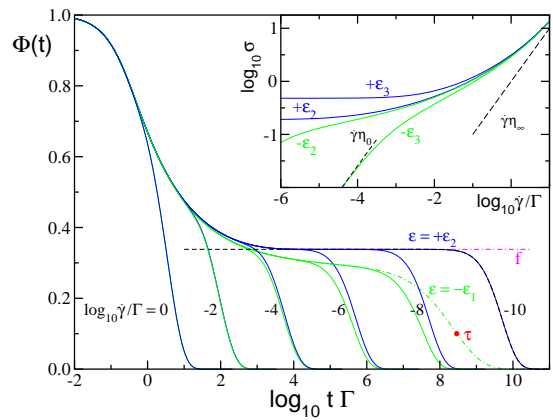}
\caption{Correlators of the schematic
  $F_{12}^{{(\gd)}}$--model as function of rescaled time
 $t\Gamma$. The curve marked with a relaxation time $\tau$ taken
 at $\Phi(\tau)=0.1$ corresponds to a fluid state without shear, 
 $\eps=-\eps_{1}=-10^{{-3.791}}$ and $\gd=0$. For increasing shear
 rates, $\log_{10}(\gd/\Gamma)=$ -8, -6, -4, -2, 0, the correlators decay
 more quickly.  The curve marked by the long time plateau value
 $f_{}$ corresponds to a glass state without shear, 
 $\eps=\eps_{2}=16 \eps_{1}$ and $\gd=0$. For nonvanishing shear rates,
$\log_{10}(\gd/\Gamma)=$ -10, -8, -6, -4, -2, 0, the correlators decay 
more quickly to
 zero. The scaling function $\phi^{+}$, whose integral gives the yield
 stress,  is shown by a dashed curve shifted onto the $\eps_{2}$ and
 $\gd/\Gamma=10^{-10}$ curve.\newline
The inset shows  flow curves of the model, viz. dimensionless stress
$\sigma$ versus $\gd/\Gamma$, for two states that would be fluid respectively
 glassy without shear;  $\eps=\pm \eps_{2}$ and
 $\eps=\pm \eps_{3}=\pm 16 \eps_{2}$. Straight lines with slope unity
 indicate the variation following the
low shear $\eta_{0}$ and high shear viscosity $\eta_{\infty}$.  
\label{bild5h} }
\end{figure}

For the parameters of the model, values studied in the literature shall
be taken \cite{Goetze84,Goetze91b,Fuchs03}. The parameter $\Gamma$ sets the time scale and determines the short
time dynamics. The bare Peclet number for the model thus is given by
Pe$_{0}=\gd/\Gamma$. Following earlier studies, the
two interaction parameters are chosen as
$v_2=v_2^c=2$ and $v_1=v_1^c+\eps/(\sqrt{v^{c}_{2}}-1)$, 
where $v_1^c=v_2^c(\sqrt{4/v_2^c}-1)\approx
0.828$. Thus, a dependence on temperature or packing fraction enters the model only via
$\eps$. A glass transition singularity lies at $\eps=0$, where the
long time limit $\Phi(t\to\infty)=f$ jumps discontinuously from zero for 
$\eps<0$
to a finite value $f\ge f_{c}=1-1/\sqrt{v^{c}_{2}}$ for
$\eps\ge0$ and $\gd=0$. The parameter $f$ plays the role of the elastic
constant $G_{\infty}$ in this model. 

Comparing the amplitude $f$ of the yielding process in the
 $F_{12}^{(\gd)}$--model with the elastic constants $G_\infty$
 calculated for hard spheres in  
Sect. II.B, one notices a quantitative difference; see Fig. \ref{bild6}. While a change of
 $\eps$ by 10 \% causes an 
increase in $G_\infty$  by a factor 8, in the schematic model $f$
 changes only by 15 \%. Fitting 
experimental flow curves of particles with strong repulsive
 interactions \cite{Fuchs04} thus the present  
$F_{12}^{(\gd)}$--model can be expected to underestimate the increase
 of the stress values in 
the glassy region.

The presence of a glassy arrested structure is equivalent to 
a frozen in part in the correlator or memory function; 
thus without shear $\Phi(t\to\infty)=f>0$ and
$m(t\to\infty)=g>0$ hold for $\eps\ge0$. With
shear a non--decaying 
part in $m(t)$ is impossible, as $m(t\gd\gg1)\le (v_1+v_2)/(\gd
t)^2$; as a consequence, also $\Phi(t)$ always decays to zero. Memory is cut 
off at long times, and \gl{sm2} gives the most simple ansatz
recovering this effect 
of shear advection in the microscopic equations, and the obviously
required symmetry in 
$\gd$.  Figure \ref{bild5h} shows for fluid and glassy, at $\gd=0$,
 solutions of the $F_{12}^{(\gd)}$--model the speed up of the
 relaxation caused by increasing 
shear rates. Integrating over the correlators, as given in \gl{sm3},
leads to the viscosity 
which consequently exhibits shear thinning; representative flow curves
are shown in the inset of 
Fig. \ref{bild5h}.

It is interesting to analyse the origin of a yield stress more closely
in the schematic model. 
This can be done most easily at the glass transition point, where the
correlator arrests at 
the (critical) value of plateau at long times, $\Phi(t\to\infty)=f_c$
at $\eps=0$ and $\gd=0$;  
this quantity may be interpreted as the elastic constant $G_{\infty}^{c}$of the
amorphous solid at melting.  
Applying shear, and assuming that the unique steady state has been
reached after waiting sufficiently, 
the schematic model predicts a finite (critical) yield stress 
$\sigma^+_c=\int_0^\infty dx \Phi_c^+(x) $, that follows from
the final decay of the (elastic) constant of the glass. 
The existence of a  yield stress thus is seen to arise from the final
relaxational process in the transient, 
which is only caused by shearing and whose relaxation time is given by
$1/|\gd|$.  
The phenomenology of the generalized Maxwell model of Sect. II.A is
thus recovered in the schematic model, and can be recognized in
Fig. \ref{bild5h}. The part frozen in without shear of magitude $f$,
decays during a process driven by and characterized by the same time
scale as the external shear rate.

\begin{acknowledgments}
We thank T. Franosch, 
J.-L. Barrat, J. Bergenholtz, L. Berthier,
A. Latz and G. Petekidis for discussions. 
M.F. thanks M.E. Cates with whom the
        theoretical approach was developed for enlightening discussions.
M. B. was supported by the DFG, SFB 481, Bayreuth.
\end{acknowledgments}


\begin{thebibliography}{10}

\bibitem{russel}
W.~B. Russel, D.~A. Saville, and W.~R. Schowalter, {\em Colloidal Dispersions}
  (Cambridge University Press, New York, 1989).

\bibitem{larson}
R.~G. Larson, {\em The Structure and Rheology of Complex Fluids} (Oxford
  University Press, New York, 1999).

\bibitem{Megen93}
W. van Megen and S.~M. Underwood, Phys. Rev. Lett. {\bf 70},  2766  (1993); {\bf 72}, 1773 (1994).

\bibitem{Megen93b}
W. van Megen and S.~M. Underwood, Phys. Rev. E {\bf 47},  248  (1993).

\bibitem{Megen94}
W. van Megen and S. Underwood, Phys. Rev. E {\bf 49},  4206  (1994).

\bibitem{Beck99}
C. Beck, W. H\"artl, and R. Hempelmann, J.~Chem.~Phys.  {\bf 111}, 8209 (1999).

\bibitem{Bartsch02} E. Bartsch, T. Eckert, C. Pies,  and H. Sillescu,
J.~Non-Cryst.~Solids {\bf 307--310}, 802 (2002).

\bibitem{Eckert03}
T. Eckert and E. Bartsch, Faraday~Discuss. {\bf 123}, 51 (2003).

\bibitem{Mason95}
T.~G. Mason and D.~A. Weitz, Phys. Rev. Lett. {\bf 75},  2770  (1995).

\bibitem{Fuchs04}
M. Fuchs and M. Ballauff, J. Chem. Phys.,  {\bf 122}, in press (2005).

\bibitem{Fuchs02}
M. Fuchs and M.~E. Cates, Phys. Rev. Lett. {\bf 89},  248304  (2002).

\bibitem{Fuchs03}
M. Fuchs and M.~E. Cates, Faraday Disc. {\bf 123},  267 (2003).

\bibitem{Fuchs03b} M. Fuchs and M. E. Cates, J. Phys.: Cond. Mat. {\bf 15}, S401 (2003).

\bibitem{Cates04}
M.~E. Cates, K. Kroy, W.~C.~K. Poon, A.~M. Puertas, and M. Fuchs, J. Phys.:
  Condens. Matter {\bf 16}, S4861 (2004).

\bibitem{Goetze91b}
W. G{\"o}tze,  in {\em Liquids, Freezing and Glass Transition}, edited by J.-P.
  Hansen, D. Levesque, and J. Zinn-Justin (North-Holland, Amsterdam, 1991), p.\
  287.

\bibitem{gs}
W. G\"otze and L. Sj\"ogren, Rep. Prog. Phys. {\bf 55},  241  (1992).

\bibitem{Goetze99}
W. G{\"o}tze, J. Phys.: Condens. Matter {\bf 11},  A1  (1999).

\bibitem{Forster75}
D. Forster, {\em Hydrodynamic Fluctuations, Broken Symmetry, and Correlation
  Functions} (W.A. Benjamin, Reading, MA, 1975).

\bibitem{Batchelor77}
G.~K. Batchelor, J. Fluid Mech {\bf 83},  97  (1977).

\bibitem{Lionberger94}
R.~A. Lionberger and W.~B. Russel, J. Rheol {\bf 38},  1885  (1994).

\bibitem{Trappe00}
V. Trappe and D. A. Weitz, Phys. Rev. Lett. {\bf 85},  449  (2000).

\bibitem{Prasad03}
V. Prasad, V. Trappe, A. D. Dinsmore, P. N. Segre, L. Cipelletti, and
 D. A. Weitz, Faraday Disc. {\bf 123},  1 (2003).

\bibitem{Brady93}
J.~F. Brady, J. Chem. Phys. {\bf 99},  567  (1993).

\bibitem{fuchsmayr}
M. Fuchs and M.~R. Mayr, Phys. Rev. E {\bf 60},  5742  (1999).

\bibitem{Franosch98}
T. Franosch, W. G{\"o}tze, M.~R. Mayr, and A.~P. Singh, J. Non-Cryst. Solids
  {\bf 235--237},  71  (1998).

\bibitem{Cates04b}
  M.E. Cates, C.B. Holmes, M. Fuchs and O. Henrich,
 in {\it Unifying Concepts in Granular Media
  and Glasses},  edited by   A. Coniglio et al.,
  (Elsevier, Amsterdam, 2004), p. 203.
  
\bibitem{Goetze84}
W. G{\"o}tze, Z. Phys. B {\bf 56},  139  (1984).


\end{thebibliography}

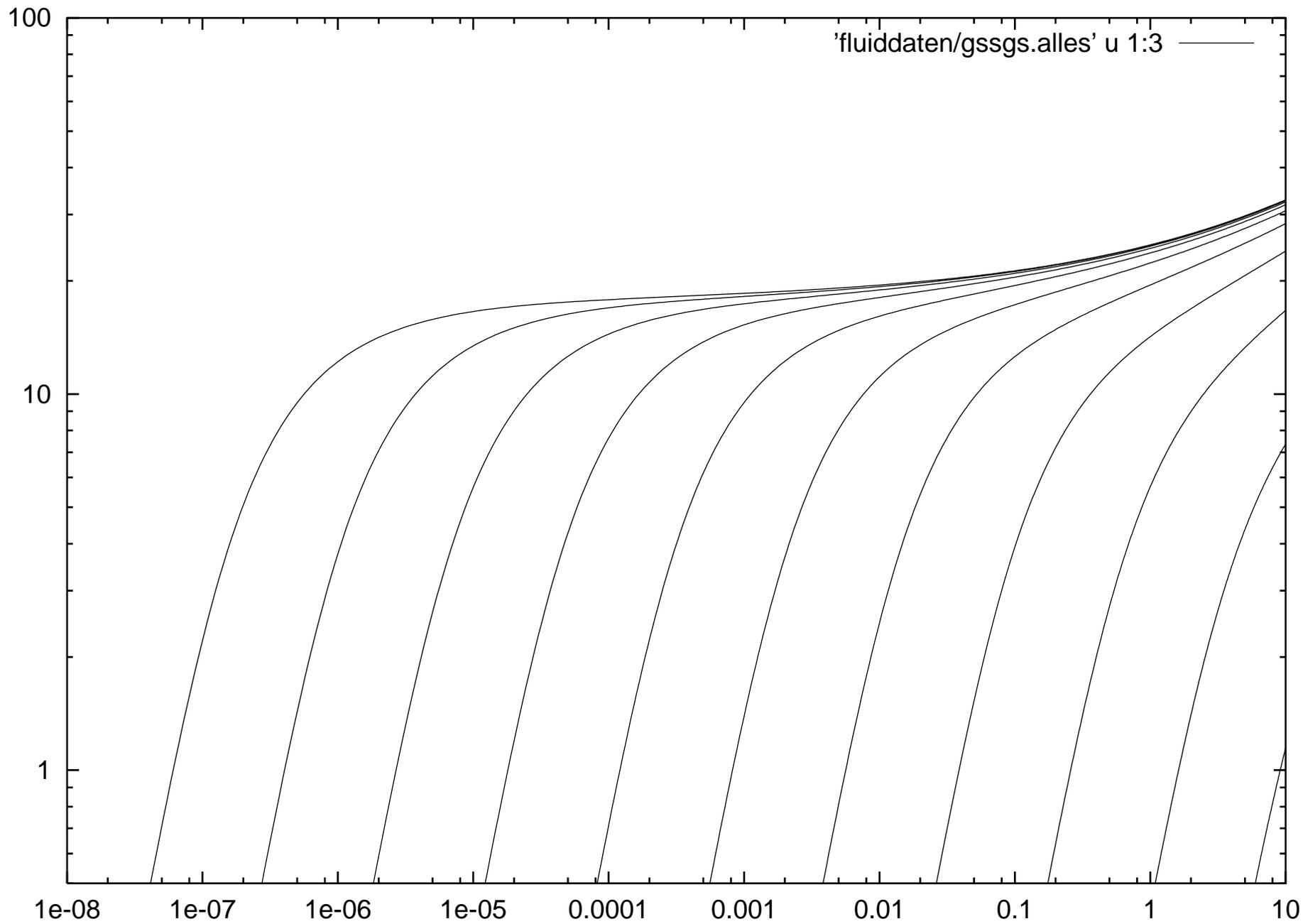
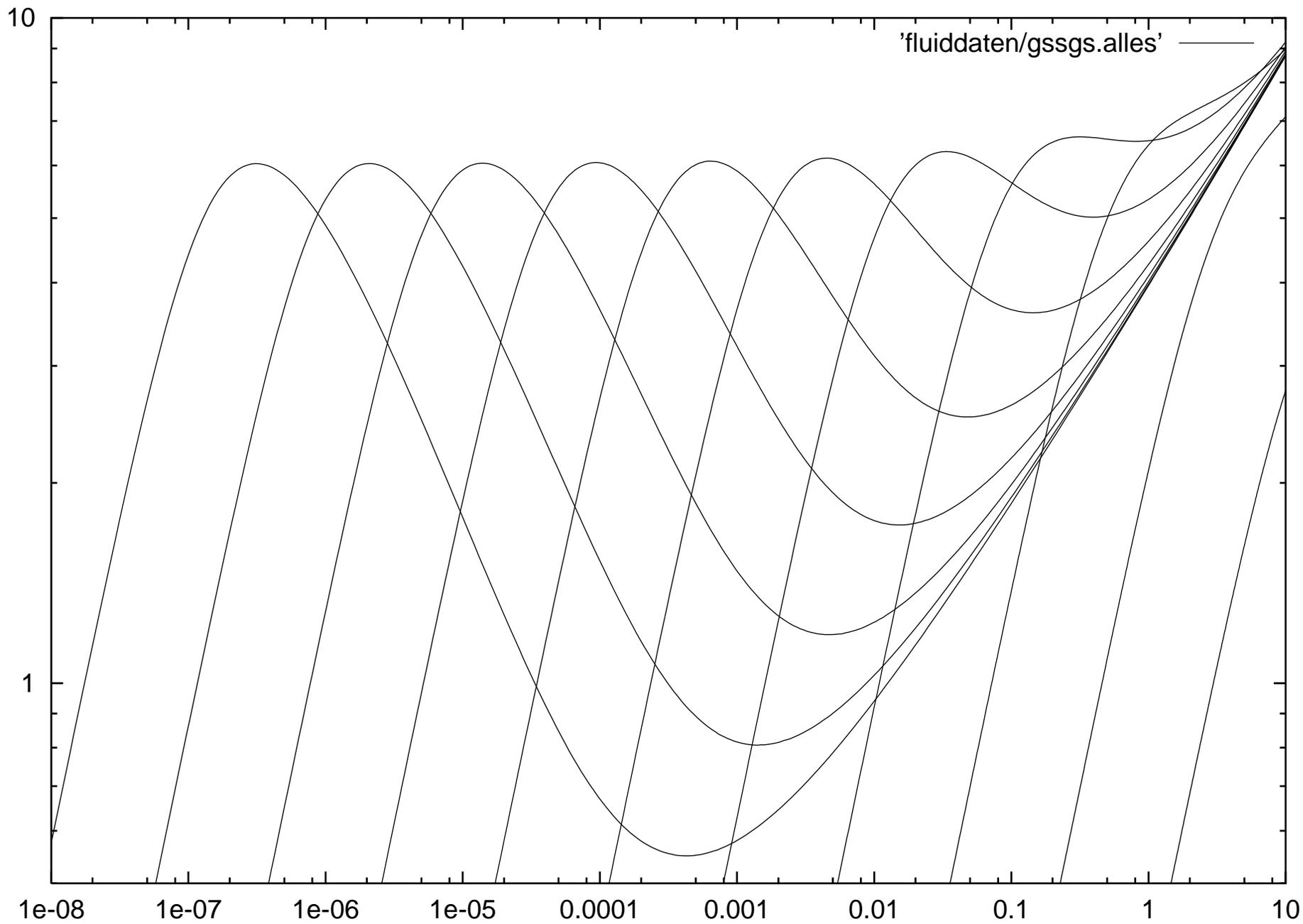
\end{document}